\documentclass[conference]{IEEEtran}
\IEEEoverridecommandlockouts
\usepackage{cite}
\usepackage{amsmath,amssymb,amsfonts}
\usepackage{algorithmic}

\usepackage{graphicx}
\graphicspath{{images/}}
\usepackage{textcomp}
\usepackage{xcolor}
\def\BibTeX{{\rm B\kern-.05em{\sc i\kern-.025em b}\kern-.08em
    T\kern-.1667em\lower.7ex\hbox{E}\kern-.125emX}}

\usepackage{balance}
\usepackage{adjustbox}
\usepackage{pifont}  %

\usepackage{siunitx}
\usepackage{hyperref}
\usepackage{lipsum}
\usepackage{subcaption}  %
\usepackage[hang]{footmisc}

\begin{document}

\title{LEAF: Simulating Large Energy-Aware \\ Fog Computing Environments
}

\author{
\IEEEauthorblockN{Philipp Wiesner and Lauritz Thamsen}
\IEEEauthorblockA{\textit{Technische Universität Berlin, Germany} \\
\{wiesner, lauritz.thamsen\}@tu-berlin.de}
}

\maketitle

\begin{abstract}
Despite constant improvements in efficiency, today's data centers and networks consume enormous amounts of energy and this demand is expected to rise even further.
An important research question is whether and how fog computing can curb this trend.
As real-life deployments of fog infrastructure are still rare, a significant part of research relies on simulations.
However, existing power models usually only target particular components such as compute nodes or battery-constrained edge devices.

Combining analytical and discrete-event modeling, we develop a holistic but granular energy consumption model that can determine the power usage of compute nodes as well as network traffic and applications over time.
Simulations can incorporate thousands of devices that execute complex application graphs on a distributed, heterogeneous, and resource-constrained infrastructure.
We evaluated our publicly available prototype LEAF within a smart city traffic scenario, demonstrating that it enables research on energy-conserving fog computing architectures and can be used to assess dynamic task placement strategies and other energy-saving mechanisms.

\end{abstract}


\section{Introduction}\label{sec:intro}

With the rise of the internet of things and the beginning of the fourth industrial revolution, many sectors and environments are expected to become interspersed with connected sensors. As these sensors produce an ever-increasing amount of data, the classical cloud computing approach of sending all information directly to a few powerful data centers, often far away from the data source, is not feasible anymore.
One attempt to solve these emerging problems is to deploy a distributed layer of compute nodes close to the edge of the network that filters, aggregates, and preprocesses the raw data generated by end devices and sensors. This new paradigm, called fog computing, promises to save large amounts of bandwidth, reduce latencies for time-sensitive applications, and increase end-user privacy as the fog nodes can take care of aggregating and anonymizing data~\cite{Dastjerdi_FogComputingHelpingIoTRealizeItsPotential_2016}. However, due to their distributed and heterogeneous nature, fog computing environments are much more challenging to manage than centralized data centers~\cite{Yousefpour_All_one_needs_to_know_2019}.

At the same time, energy consumption of information and communications technology is already accounting for more than \SI{10}{\percent} of global energy consumption and is expected to exceed the \SI{20}{\percent} mark by 2030~\cite{Nature_Electricity_Forecast_2018}. An open research question is whether fog computing will be able to reduce the overall energy consumption of future computing infrastructure or if it may even be responsible for a further increase\mbox{\cite{Jalali_FogComputingMayHelp_2016, Dastjerdi_FogComputingHelpingIoTRealizeItsPotential_2016, Noor_MCCChallengesAndFutureResearchDirections_2018}}. Although the general trend of concentrating computing resources into hyper-scale data centers continues~\cite{DataCenterFrontier_HyperscaleDataCentersReport_2019}, research has shown that distributed architectures could save between \SI{14}{\percent} to more than \SI{80}{\percent} of total energy consumption compared to fully centralized architectures\mbox{\cite{Ahvar_AnalyticalCloudFogEdge_2019, Yan_Modeling_Consumption_Mobile_Network_2019}}.

A major part of today's fog computing research is centered around modeling and simulations. Nevertheless, to the best of our knowledge, there exists no tool that is suitable to explore the overall power usage of large-scale fog deployments and enables research on energy-conserving architectures and algorithms in this domain. Major existing fog computing simulators provide no energy modeling at all~\cite{Sonmez_EdgeCloudSim_2018, Zeng_IOTSim_2016, Lera_YAFS_2019} or only focus on specific components of fog infrastructures~\cite{Gupta_iFogSim_2017, Mechalikh_PureEdgeSim_2019}. Furthermore, many simulators fail to realistically model fog environments by leaving out substantial aspects like mobility~\cite{Gupta_iFogSim_2017, Zeng_IOTSim_2016}. Existing analytical approaches on the other hand provide only a high level of abstraction and cannot represent change over time, prohibiting the modeling of mobility and evaluation of online decision making algorithms~\cite{Sarkar_TheoreticalModellingOfFogComputing_2016, Jalali_FogComputingMayHelp_2016, Ahvar_AnalyticalCloudFogEdge_2019}.

We present LEAF\footnote{Java implementation available at \url{https://github.com/dos-group/leaf-java}\linebreak Python implementation available at \url{https://github.com/dos-group/leaf}}, a new simulator for modeling \underline{L}arge \underline{E}nergy-\underline{A}ware \underline{F}og computing environments.
In contrast to the related work, our model was designed with the following four requirements in mind:

\begin{enumerate}
  \item[(1)] \emph{Realistic Fog Computing Environment}: Enable the simulation of applications with complex task graphs in distributed, heterogeneous, and resource-constrained fog computing environments. Nodes can be mobile and they can join or leave the topology at any time.
  
  \item[(2)] \emph{Holistic Energy Consumption Model}: Enable the separate assessment of power usage of compute nodes, network traffic, and applications, allowing for the creation of load profiles for individual parts of fog computing environments.
  
  \item[(3)] \emph{Energy-Aware Online Decision Making}: Enable research on task placement strategies, routing policies, and energy-saving mechanisms that dynamically adapt their behavior based on the power usage of infrastructure and applications.
  
  \item[(4)] \emph{Performance and Scalability}: Enable the simulation of complex scenarios with thousands of devices and applications at least two orders of magnitude faster than real-time on commodity hardware.
\end{enumerate}

Our model combines analytical and discrete-event modeling to enable easy-to-analyze large-scale experiments.
Specifically, we model infrastructures and applications as variable graphs that can change throughout the simulation and do not rely on representing individual packets.
This way, the model enables research on novel task placement strategies or energy-saving mechanisms that must be evaluated in large settings.
We expect LEAF to help both researchers and practitioners assess the power usage characteristics of different cloud, fog, and edge computing architectures, which can lead to more well-founded decisions when planning future infrastructures.

The remainder of the paper is structured as follows.
Section~\ref{sec:RELATED_WORK} discusses the related work. 
Section~\ref{sec:SIMULATION_MODEL} introduces our infrastructure and application model for fog computing environments.
Section~\ref{sec:POWER_MODELS} explains the different kinds of power models that can be applied.
Section~\ref{sec:EVALUATION} evaluates our model on a smart city scenario.
Section~\ref{sec:CONCLUSION} concludes the paper.

\section{Related Work}\label{sec:RELATED_WORK}

Energy-efficient data centers have been a research area for years as electricity costs account for a significant part of a data center's operating expenditure.
In fog or edge computing, however, most research on power usage only concerns specific components such as battery-constrained edge devices~\cite{Mechalikh_PureEdgeSim_2019} or fog nodes~\cite{Gupta_iFogSim_2017}.
The bigger picture, namely the potential impact of fog computing on the overall energy consumption of information and communications technology, has retrieved comparably little attention.

One approach to assess the estimated consumption of fog computing environments is analytical modeling. These models are usually suitable for examining large-scale scenarios since they are fast to execute and convenient to experiment with.
Sarkar et al.~\cite{Sarkar_TheoreticalModellingOfFogComputing_2016} compare service latency and energy consumption of fog and cloud computing by performing a high-level examination of different architectures. The authors focus on infrastructure only and incorporate no concept for applications. Furthermore, they do not distinguish between static and dynamic energy consumption, which limits the informative value of the analyses.
Jalali et al.~\cite{Jalali_FogComputingMayHelp_2016} model energy usage by using a flow-based model for highly shared equipment such as core routers and switches, and a time-based model for end-user equipment that is usually not shared among many clients. Their findings show that the power impact of fog computing depends on many factors such as the type of attached access network. However their scenario is a relatively specific use case and not suitable as a general purpose model.
Ahvar et al.~\cite{Ahvar_AnalyticalCloudFogEdge_2019} propose a model that differentiates between static and dynamic power usage and incorporates the consumption of network devices and cooling. Even though their model is relatively comprehensive, it incorporates several critical simplifications such as an even distribution of VMs among data centers and the assumption that all fog nodes connect to the same number of devices. Moreover, there exists no concept of applications or tasks, nor is there a way to calculate the relative power requirements of VMs.  %
All presented analytical approaches suffer from the fact that they model at a high level of abstraction and, for example, make the invalid assumption that fog nodes are homogeneous~\cite{Sarkar_TheoreticalModellingOfFogComputing_2016, Ahvar_AnalyticalCloudFogEdge_2019}.
Moreover, they cannot represent change over time, prohibiting the assessment of mobility and online decision making algorithms.

Besides analytical models, a large number of discrete-event simulators has been developed in recent years.
Although there exist cloud computing simulators that feature comprehensive power models like GreenCloud~\cite{Kliazovich_GreenCloud_2010}, these tools are not designed to model highly heterogeneous and variable environments like those encountered in fog computing.

Many popular simulators in the domain of fog and edge computing such as EdgeCloudSim~\cite{Sonmez_EdgeCloudSim_2018} or IOTSim~\cite{Zeng_IOTSim_2016} comprise no power modeling at all.
Others, like YAFS~\cite{Lera_YAFS_2019}, allow users to specify custom attributes related to node power usage but do not feature any power modeling at runtime.
The simulators that do are either overly simplistic or focus only on specific components of the infrastructure.
For example, iFogSim~\cite{Gupta_iFogSim_2017} inherits some power modeling capabilities from CloudSim and can model the power consumption of data centers and fog nodes. Nevertheless, there is currently no way to simulate the power usage of edge devices, networks, or applications. 
PureEdgeSim~\cite{Mechalikh_PureEdgeSim_2019} features a comprehensive energy consumption model for edge devices, including Wi-Fi traffic, to simulate battery-constrained devices. However, the consumption of data centers, fog nodes or wide area network (WAN) traffic cannot be modeled.
FogNetSim++~\cite{Qayyum_FogNetSim++_2018} is the only fog computing simulator to our knowledge that features an energy model for applications.
Albeit the focus on performance, the detailed modeling of network traffic on a per packet basis inevitably results in worse scalability than our model. Furthermore, due to our analytical approach, it is a lot easier to set up experiments in LEAF, explore different configurations, and analyze results.

\section{Fog Environment Model}\label{sec:SIMULATION_MODEL}

\begin{figure}[b]
    \centering
    \includegraphics[width=1\linewidth, trim=0cm 0cm 0cm 0.2cm, clip]{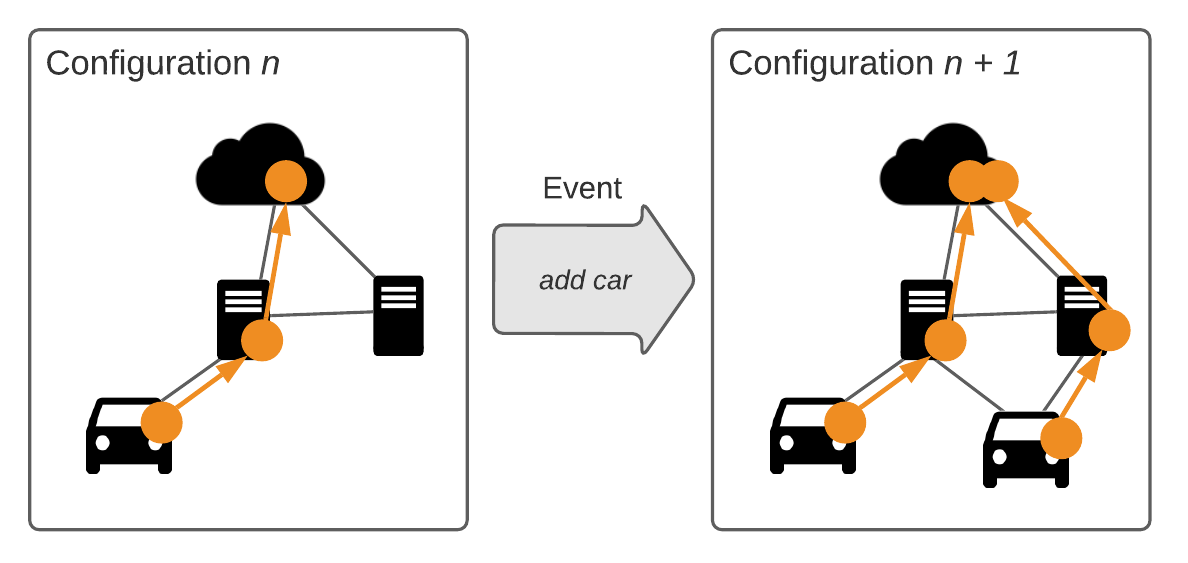}
	\caption{LEAF uses discrete events to update the state (configuration) of the simulation. In this example, an additional compute node (a car) was added to the infrastructure graph and a new application was placed on this infrastructure.}
    	\label{fig:DES}
\end{figure}

The way LEAF models fog computing environments is inspired by analytical modeling. 
Infrastructure and applications are represented as graphs and power models as functions that operate on these graphs.
Together, these components define the state of the simulation at a certain time step, which we call a \emph{configuration}.
In contrast to purely analytical models, the state of network and applications can change over time through the use of events, as illustrated in Figure~\ref{fig:DES}.
LEAF uses events only in two cases: To read and to update configurations.
Read events can, for example, be emitted by an algorithm that performs power measurements at a certain time step to adapt its behavior.
Examples of update events are changes to the current application placements or the infrastructure graph like adapting the latency of a network link or adding a new compute node.
LEAF does not use events to simulate individual network packets.

Every configuration contains all information necessary for a deeper analysis and can be inspected without any knowledge about future or past events.
Modeling fog computing environments in this way enables easy-to-analyze experiments that can execute scenarios with thousands of devices and applications.
Yet, our approach also has decisive advantages over purely analytical models: First, by allowing the model to change over time, heterogeneous environments with dynamically changing network topologies, mobility of edge devices, and varying workloads can be represented. Second, LEAF allows for online decision making, enabling research on dynamic energy-aware task placement strategies, scheduling algorithms, or traffic routing policies.

\subsection{Infrastructure Model}

\begin{figure}
    \centering
    \includegraphics[width=1\linewidth]{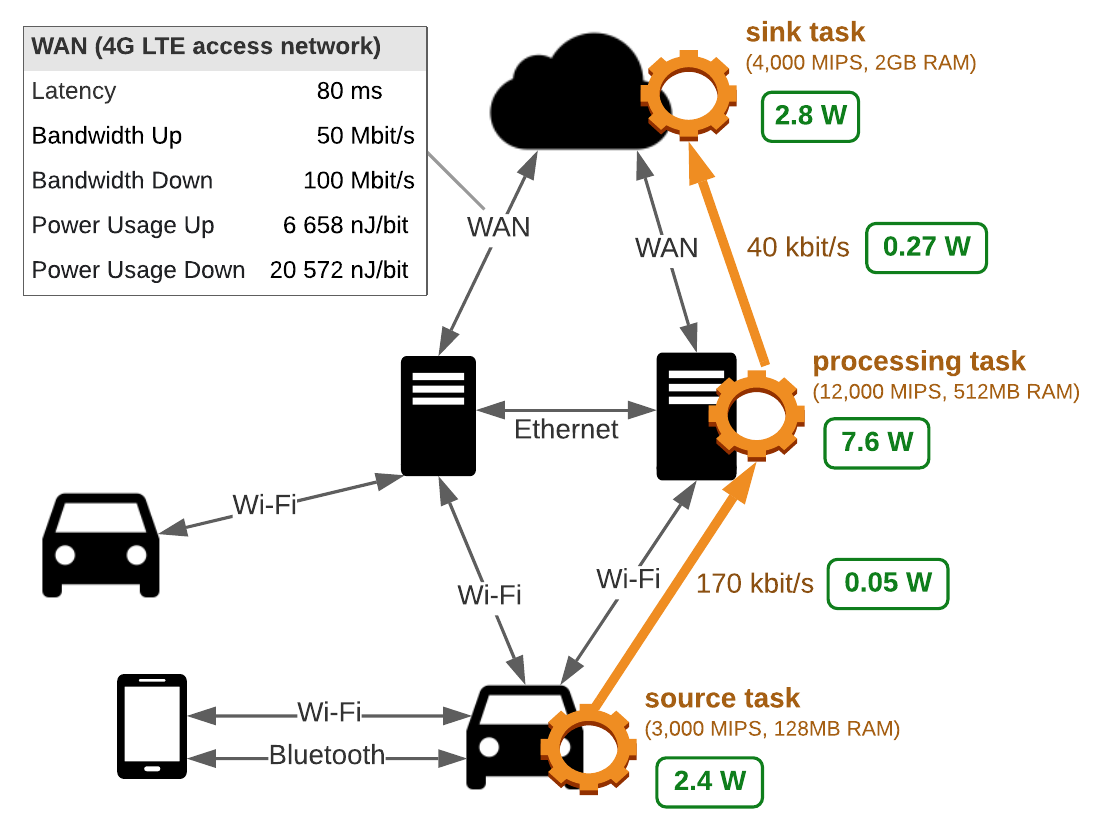}
	\caption{Infrastructure graph with resource constraints and power usage characteristics (black), a placed application with resource requirements (orange), and its resulting power usage on the infrastructure components (green). In this configuration, the application's combined power usage is \SI{13.12}{W}.}
    	\label{fig:model}
\end{figure}

As depicted in Figure~\ref{fig:model}, the network and computing infrastructure are represented as a weighted, directed multigraph $I = (N, L),$ where
\begin{itemize}
    \item[$N$] is the set of compute nodes. A compute node $N^{i} \in N$ may be resource-constrained, can have a location, and may be mobile. It can represent an entire cloud data center as well as mobile sensors with limited or no computing capacity.
    \item[$L$] is the set of network links between nodes. A network link $L^{i} \in L$ can represent individual wired or wireless connections as well as entire network routes such as WAN connections. Network links can be constrained by bandwidth and may have additional properties like latency that users can utilize for implementing routing policies. Edges are directed to allow modeling different constraints and power models for uplink and downlink. Multiple edges between nodes are allowed and can represent different available link types such as Wi-Fi and Bluetooth.
    \end{itemize}

The infrastructure graph can change during the simulation, as nodes are free to move around and can join or leave the network. Furthermore, any property, like the latency of a link, can change.

\subsection{Application Model}

LEAF considers all applications to be streaming applications and, hence, regards all communication between different application tasks as continuous data flows. Applications are represented by directed acyclic graphs $A = (T, F),$
where $T$ is the set of tasks in the application, and $F$ is the set of data flows between tasks. Tasks, as well as data flows, have certain resource requirements. Applications can be placed on the fog by mapping their tasks to compute nodes, and their data flows to network links. Application placements are described in more detail below.

A task $T^{i} \in T$ demands certain resources from its compute node. A compute node can host as many tasks as its resources allow. There exist three different kinds of tasks that are responsible for emitting, processing, and retrieving data:

\begin{enumerate}
    \item \emph{Source tasks} are bound to a specific compute node, for example, a sensor-equipped or mobile device. They are constantly emitting data and have at least one outgoing edge and no incoming edges.
    \item \emph{Processing tasks} can be freely placed on any compute node $N^{i} \in N$ that fulfills their resource requirements. Processing tasks have at least one incoming edge and at least one outgoing edge.
    \item \emph{Sink tasks} are again bound to a specific compute node and have at least one incoming edge and no outgoing edges. Examples of data sinks are cloud storages or end-user terminals.
\end{enumerate}

A data flow $F^{i} \in F$ represents a continuous stream of data between two tasks, stated in \si{bits/s}. Like any parameter, the data rate can change during the simulation. A network link can ``host'' as many data streams as its bandwidth allows. Due to its abstract network model, LEAF is not suitable to explicitly model complex effects resulting from network congestion like queuing delay or packet drops.

When placing an application, its tasks have to be mapped onto compute nodes, which is illustrated in orange in Figure~\ref{fig:model}. Since the assignment of source and sink tasks is fixed, task placement strategies have to determine the placement of processing tasks. After the placement, a user-defined routing policy searches for an optimal path in the network between every pair of connected tasks. If there is no node that meets the resource requirements or if no sufficient path can be found, the placement failed. %

Typically, processing tasks aggregate and filter the data stream in a way that outgoing data flows require less bandwidth than incoming data flows. Consequently, the placement of processing tasks close to the source tasks generally reduces network usage and, thus, energy consumption. Finding optimal placements in dynamic, resource-constrained, distributed, and heterogeneous environments is a complicated problem and a focus of research~\cite{Brogi_HowToPlaceYourAppsInTheFog_2019}. LEAF enables application placement strategies that take the infrastructure's energy consumption into account.

\section{Power Models}\label{sec:POWER_MODELS}

We model infrastructure power usage by assigning each compute node and network link its own power model. This enables users to determine the power usage of data centers, edge devices, and network links individually. A power model $P(t)$ is a function that returns the current power usage of its corresponding entity at any point in time $t$ during the simulation. Power usage is reported as the sum $P_{static} + P_{dyn}$, where $P_{static}$ is the static (load-independent) power usage and $P_{dyn}$ the dynamic (load-dependent) fraction. Power models may maintain a state which can be used to implement energy-saving mechanisms (see Section \ref{sec:energy-saving-mechanisms}).

To create load profiles of infrastructure components, their power models can be called through periodic events. Additionally, other simulated components can call power models at runtime to adapt their behavior or update their state. Examples of such components are energy-aware task placement strategies or hardware like batteries that update the state of charge.

\subsection{Linear and Non-Linear Power Models}

Research shows that the power usage of infrastructure can often be modeled with sufficient accuracy as a linear function that is only based on the entity's current load $C(t)$~\cite{Barroso_Hoelzle_EnergyProportionalComputing_2007, Hinton_EnergyConsumptionModellingOfOpticalNetworks_2015}. %
Linear power models are beneficial since they are computationally efficient and easy to comprehend. They are described~as
$$P(t) = P_{static} + C(t)\, \sigma,$$
where $C(t)\, \sigma$ represents $P_{dyn}$. The variable $\sigma$ determines the incremental energy per load. For example, for network equipment, $\sigma$ represents the energy consumed per bit of transferred data (\si{J}/\si{bit}).
When modeling resource-constrained compute nodes or network links that have a maximum load $C_{max}$ with power usage $P_{max}$, $\sigma$ can be expressed as:
$$\sigma = \frac{P_{max} - P_{static}}{C_{max}}$$

An example for more complex entities, where a simple, linear function may not model the power usage well, are compute nodes representing entire data centers. Data centers incorporate several different hosts and require additional power for operational overhead like cooling and lighting. An exemplary approach for modeling this kind of compute node is to define a power model $P_{H_{i}}$ for each host $H_{i}$ and to aggregate their results to $P_{dc}$. To account for operational overhead, we can multiply the result with the data center's power usage effectiveness (PUE):
$$P_{dc}(t) = \si{PUE} \cdot \sum_{H_{i} \in H} P_{H_{i}}(t)$$

Compute nodes and network links can also apply non-linear power models and power models with multiple inputs to improve the accuracy of the simulation. 
An example is the power model for wireless network links of~\cite{Heinzelman_WirelessMicrosensorNetworks_2000}, which, in addition to the load, also takes the distance of communicating devices into account.
However, more complex power models will lead to more computationally intensive simulations, hence longer runtimes.

\subsection{Power Models for Shared Resources}

When describing the power consumption of shared resources such as cloud data centers, mobile base stations, or entire WAN links, it is not meaningful to specify static power usage. For example, it is impossible to know at which point an externally operated data center will need to switch on or off a new physical machine. Many of these resources are assumed to scale on demand, hence defining a $P_{max}$ is not even possible. For these reasons, $P_{static} = 0$ for power models of shared infrastructure. Users must estimate a meaningful $\sigma$ that directly incorporates a fraction of the static consumption. 

\begin{figure}[b]
    \centering
    \hspace*{-.6cm}\includegraphics[width=0.8\columnwidth]{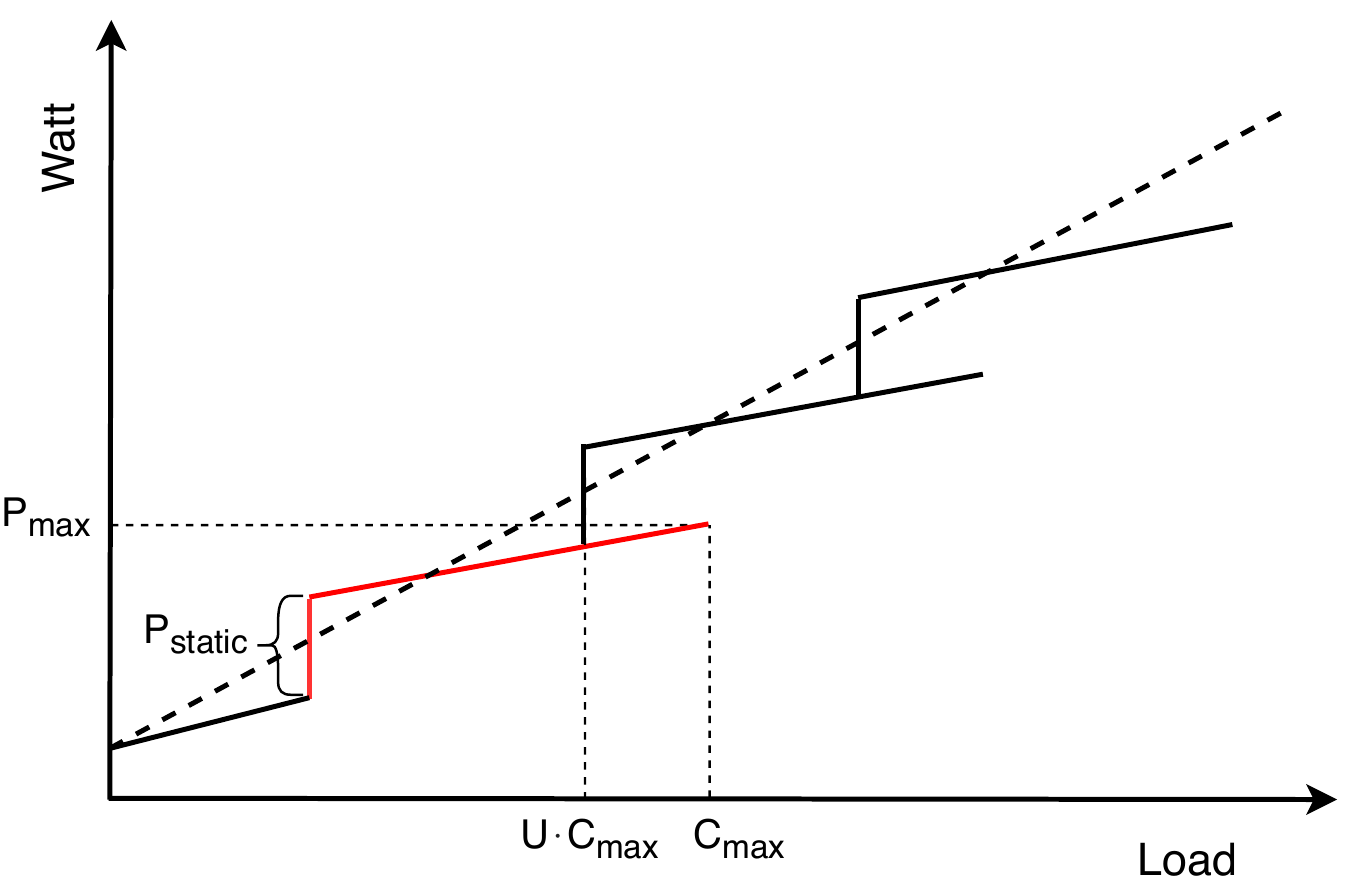}
    \caption{Incremental performance per watt of scalable compute nodes or network links: The red section describes a sub-component such as a data center host, which is only utilized until $U \cdot C_{max}$ of its full capacity. Adapted from \cite{Jalali_FogComputingMayHelp_2016}.}
    \label{fig:staircase}
\end{figure}

By modeling shared resources without static power usage, we implicitly assume this infrastructure to be energy-proportional. Nevertheless, there are ways to incorporate inefficiencies into the parameters. In practice, servers as well as routers are rarely used to full capacity but are over-provisioned to ensure performance and availability. Equivalently to the energy model proposed by \cite{Jalali_FogComputingMayHelp_2016}, users can estimate and incorporate an operational capacity fraction $U$ into a power model. The effect of $U$ is illustrated in the staircase curve depicted in Figure~\ref{fig:staircase}.

\subsection{Power Models for Network Links}

Although we assign power models directly to edges in the infrastructure graph, in reality, energy is not consumed by the cable or wireless connection itself but by networking equipment such as routers or wireless transmitters.
Network link power models hence represent the aggregated power usage of all networking equipment involved to transfer data from one compute node to another, and $\sigma$ describes the incremental energy per bit (\si{J/bit}).
Static power usage should be attributed to the adjacent compute nodes, to assure it correctly drops to zero when the node is shut off.
Intermediate routing devices are considered shared infrastructure and their consumption should be incorporated directly into the network link's $\sigma$. Table~\ref{table:WAN} depicts an exemplary parameterization of a WAN link between a mobile device and a cloud server.

\begin{table}
\small
\centering
\caption{Incremental energy per bit (\si{\nano J/bit}) of networking equipment adding up to an exemplary WAN link. Cisco hardware power consumption was determined with the Cisco Power Calculator\protect\footnotemark.}
\begin{tabular}{|l|r|r|}
\hline
Type               & $\sigma_{up}$ & $\sigma_{down}$ \\
\hline
4G LTE module of mobile device~\cite{Huang_CloseExaminationOfPerformanceAndPowerCharacteristicsOf4GLTENetworks_2012} & 438.4 & 52.0 \\
4G LTE access network~\cite{Vishwanath_EnergyConsumptionOfInteractiveCloudBasedDocumentProcessingApplications_2013} & \SI{6200.0}{} & \SI{20500.0}{} \\
Edge router (Cisco ASR 9010)      & 5.9 & 5.9          \\
Core routers (5x Cisco CRS-3)     & 13.5 & 13.5   \\
DC switches (2x Cisco Nexus 9500) & 0.4 & 0.4 \\
\hline
WAN link & \SI{6658.2}{} & \SI{20571.8}{}  \\
\hline
\end{tabular}
\label{table:WAN}
\end{table}

\footnotetext{\url{http://tools.cisco.com/cpc}, accessed 2020-10-17}

For wireless connections, power usage characteristics of uplink and downlink usually differ~\cite{Huang_CloseExaminationOfPerformanceAndPowerCharacteristicsOf4GLTENetworks_2012, Vishwanath_EnergyConsumptionOfInteractiveCloudBasedDocumentProcessingApplications_2013}.
Since infrastructure is represented by a directed multigraph in LEAF, it allows modeling asymmetrical power usage characteristics by assigning different power models to the respective directions.

\subsection{Energy-saving Mechanisms}\label{sec:energy-saving-mechanisms}

Energy-saving mechanisms aim to reduce static power usage of hardware during periods of low utilization by throttling or powering off parts of the system. They can significantly influence overall energy consumption by pushing infrastructure towards operating energy-proportionally.
Such load profiles can be modeled in LEAF by making $P_{static}$, which usually describes the load-independent power usage, indirectly load-dependent. For example, a common strategy in data centers is to consolidate workload on a few hosts and to power off the remaining hosts. $P_{static}$ of such a data center could be modeled as a step function, similar to the example in Figure~\ref{fig:staircase}.

Furthermore, to evaluate the impact of even more complex energy-saving mechanisms, users can make use of the fact that power models can maintain state. For example, \cite{Guerout_DVFS_Simulation_2013} presented a precise model for dynamic voltage and frequency scaling (DVFS), which requires the current and past utilization of the CPU to implement different strategies. LEAF's power models can store past utilization states of their assigned entity and take them into account when computing the current power usage.

\subsection{Tracing Back Power Usage to Applications}

To support the implementation of energy-conserving application placement strategies, LEAF allows tracing back the infrastructure's power usage to the responsible applications.
The power consumption of an application is defined as the sum of all power usage that its tasks and data flows cause on their allocated resources, see Figure \ref{fig:model}. 
Depending on the scenario users may decide to only attribute dynamic power usage, or dynamic and static power usage. For example, if the underlying resource is powered on, regardless of its utilization, the static energy consumption does not have to be attributed to any applications since they have no influence on it. However, if the resource can be powered off when it is idle, the placed resources actually \emph{cause} the energy usage. In this case, the static energy should be attributed to all tasks that currently run on the node depending on their fraction of the total load.

\section{Evaluation}\label{sec:EVALUATION}

We created two open source implementations of LEAF in Java and Python. The respective GitHub repositories are referenced in Section~\ref{sec:intro}. The following evaluation uses the Java implementation, based on the CloudSim Plus~\cite{Filho_CloudSimPlus_2017} simulator, and simulates different fog architectures and application placement strategies in a smart city traffic scenario.

\subsection{Experimental Setup}

\begin{figure}[b]
    \includegraphics[width=1\columnwidth, trim=1.1cm 1.5cm 2cm 2.1cm, clip]{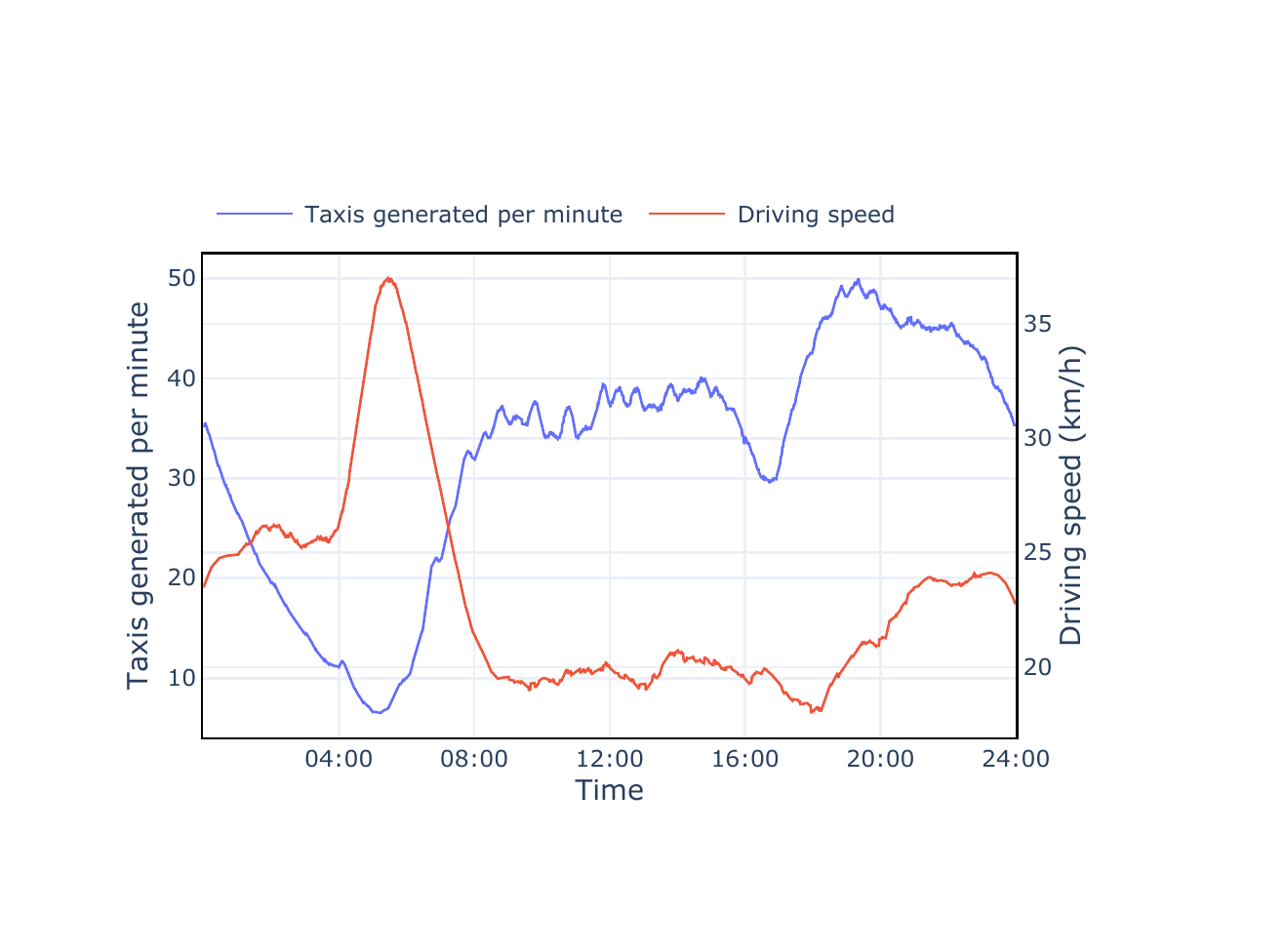}
    \caption{Number of taxis generated per minute vs. average driving speed.}
    \label{fig:taxis}
\end{figure}

The urban environment of the experimental setup is inspired by Manhattan, namely a rectangular street grid with blocks of \SI{274}{m} by \SI{80}{m}. It covers an area containing 16 crossings, each equipped with a smart traffic light (STL) that optionally has a fog node attached.
Within this city center, we simulate 24 hours of taxi traffic with time steps of one second. Quantity and speed of taxis were modeled according to a dataset published for the 2015 DEBS Grand Challenge competition\footnote{\url{http://www.debs2015.org/call-grand-challenge.html}, accessed 2020-12-15}, which contains information on approximately 173 million taxi trips that took place in New York in 2013. Figure~\ref{fig:taxis} shows the average number and speed of taxis generated per minute.

The infrastructure graph consists of four kinds of compute nodes: The cloud, fog nodes, STLs, and taxis.
Taxis are connected via Wi-Fi, e.g. \emph{IEEE 802.11p}, to nearby STLs. The mapping of data flows to network edges is updated periodically as taxis move around. STLs communicate with other close-by STLs via Wi-Fi too, effectively forming a mesh network. Additionally, they are equipped with 4G LTE modules connecting them to the cloud node via WAN.

Table~\ref{table:parameterization} shows the infrastructure parameterization. Only cloud and fog nodes have computing capacities stated in millions of instructions per second (MIPS).
Since the load on STL and taxi nodes is the same in all experiments, their power usage is not modeled in this evaluation.
All other compute nodes and network links have linear power models. The cloud is considered shared infrastructure and has no static but high dynamic power usage. Energy per bit for WAN links was configured as the example shown in Table~\ref{table:WAN}. Direct Wi-Fi communication between STLs consumes less energy than communication with taxis because they always have a direct line of sight and deploy energy-efficient, high-throughput access points.

\begin{table}
\small
\centering
\caption{Smart city infrastructure parameterization.}
\medskip
\begin{tabular}{|l|r|r|r|}
\hline
        & Max load & $P_{static}$                   & $\sigma$                    \\
\hline
Cloud   & $\infty$ & -           & \SI{700}{\micro W/MIPS} \\
Fog node & \SI{400000}{MIPS} & \SI{100}{W} & \SI{350}{\micro W/MIPS} \\
\hline
WAN \textsubscript{STL $\rightarrow$ Cloud} & \SI{50}{Mbit/s} & - & \SI{6658}{\nano J/bit} \\
WAN \textsubscript{Cloud $\rightarrow$ STL} & \SI{100}{Mbit/s} & - & \SI{20572}{\nano J/bit} \\
Wi-Fi \textsubscript{Taxi $\rightarrow$ STL}  & \SI{1.3}{Gbit/s} & - & \SI{300}{\nano J/bit} \\
Wi-Fi \textsubscript{STL $\rightarrow$ STL} & \SI{1.3}{Gbit/s} & - & \SI{100}{\nano J/bit} \\
\hline
\end{tabular}
\label{table:parameterization}
\end{table}

We simulate a smart traffic system comprising two kinds of applications:

\begin{LaTeXdescription}
    \item[CCTV] applications process data recorded by cameras deployed at STLs. The source task, located at the STL, sends \SI{10}{Mbit/s} of video data to a processing task that requires \SI{30000}{MIPS} and is responsible for traffic monitoring, enforcing traffic laws and automatic incident detection. The \SI{200}{kbit/s} of resulting data are sent to the sink task located in the cloud for further analysis and storage. 16 of these applications are running in our scenario, one for each STL.
    \item[V2I] applications are vehicle-to-infrastructure, smart traffic management applications that control traffic lights to ensure maximum throughput of public transport. Each taxi on the map streams \SI{100}{kbit/s} of sensor data to a processing task which requires \SI{7000}{MIPS} and forwards \SI{50}{kbit/s} to all STLs on the way of the taxi. The number of applications running depends on the number of taxis on the map.
\end{LaTeXdescription}

\subsection{Experiments}

\begin{figure}[h]
    \centering
    \includegraphics[width=1\columnwidth, trim=1cm 1.5cm 1.8cm 2cm, clip]{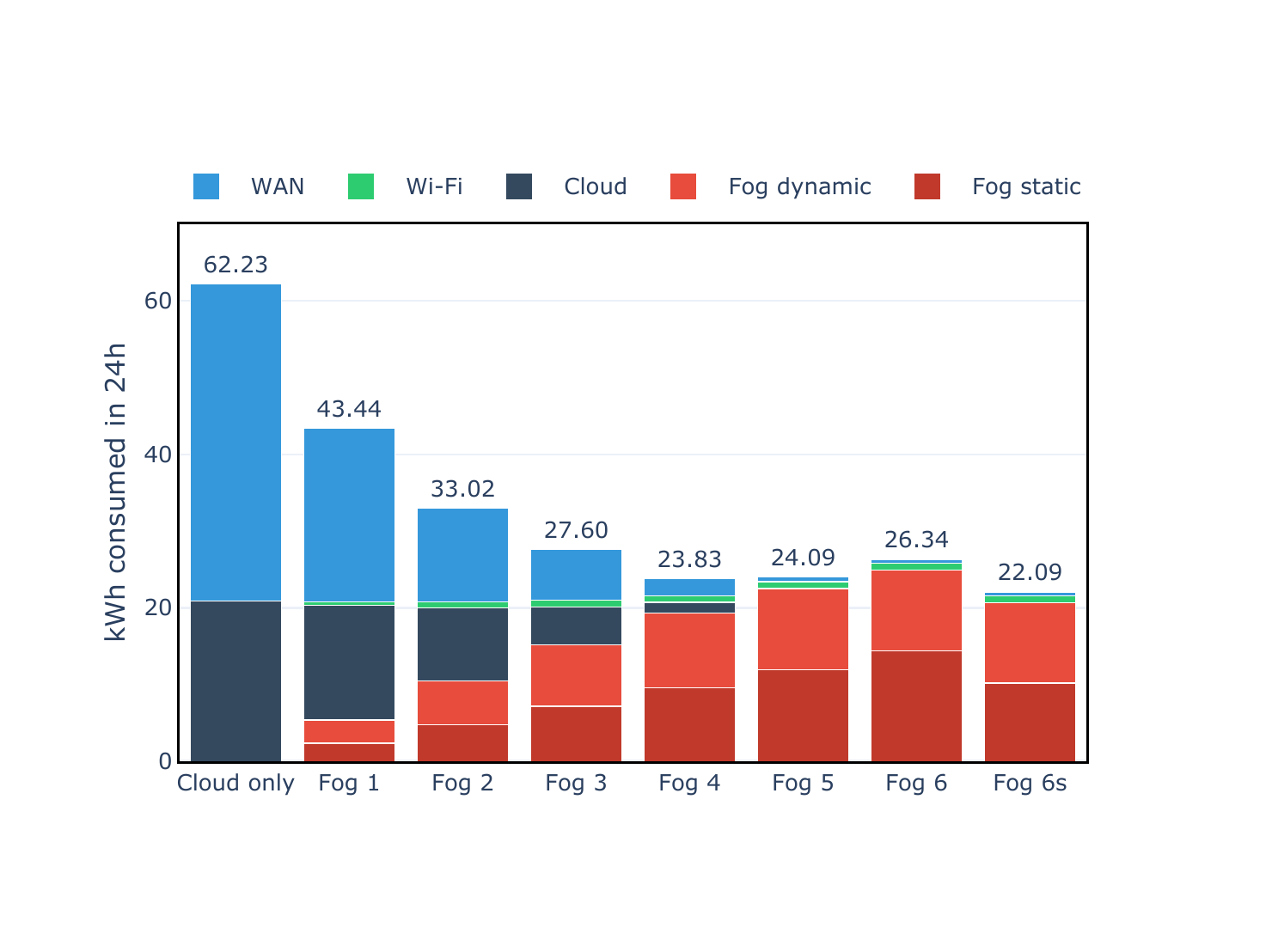}
    \caption{Aggregated energy consumption of infrastructure components of all conducted experiments.}
    \label{fig:evaluation_results}
\end{figure}

We conducted eight different experiments to demonstrate that our model enables research on energy-efficient fog architectures, task placement strategies and energy-saving mechanisms. Figure~\ref{fig:evaluation_results} provides an overview of the results. The experiments are explained and analyzed below.

\begin{figure*}[h]
    \centering
    \includegraphics[width=1\textwidth, trim=1cm 1.4cm 1.8cm 1.9cm, clip]{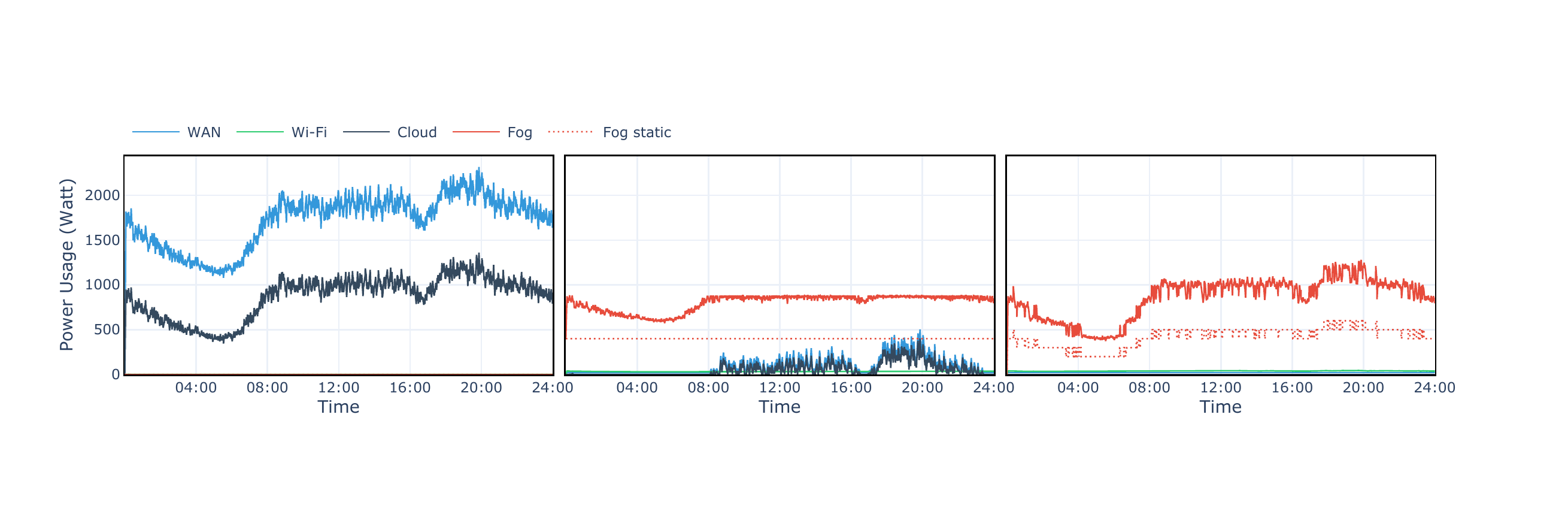}
	
	\vspace{-0.3cm}\begin{minipage}[t]{.33\linewidth}
		\centering
		\subcaption{\emph{Cloud Only} infrastructure power usage}\label{fig:cloud_infrastructure}
	\end{minipage}
	\begin{minipage}[t]{.31\linewidth}
		\centering
		\subcaption{\emph{Fog4} infrastructure power usage}\label{fig:fog4_infrastructure}
	\end{minipage}
	\begin{minipage}[t]{.3\linewidth}
		\centering
		\subcaption{\emph{Fog6s} infrastructure power usage}\label{fig:fog6s_infrastructure}
	\end{minipage}
	
    \includegraphics[width=1\textwidth, trim=1cm 1.4cm 1.8cm 1.7cm, clip]{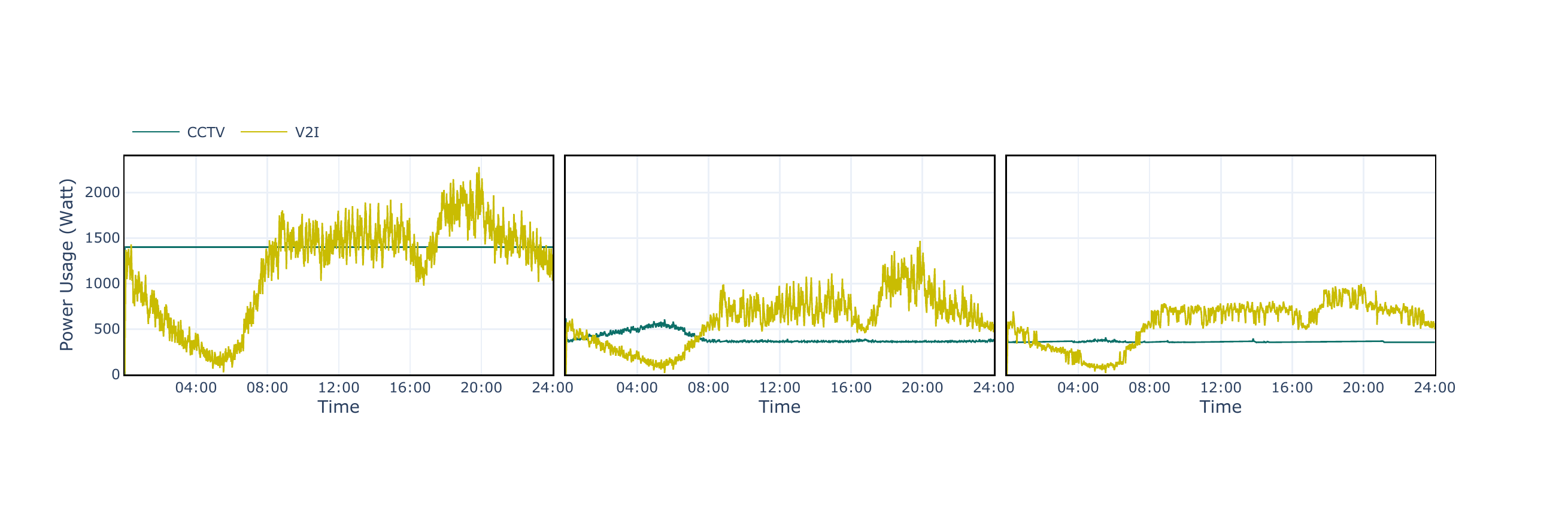}   
	
	\vspace{-0.3cm}\begin{minipage}[t]{.33\linewidth}
		\centering
		\subcaption{\emph{Cloud Only} applications power usage}\label{fig:cloud_applications}
	\end{minipage}
	\begin{minipage}[t]{.31\linewidth}
		\centering
		\subcaption{\emph{Fog4} applications power usage}\label{fig:fog4_applications}
	\end{minipage}
	\begin{minipage}[t]{.3\linewidth}
		\centering
		\subcaption{\emph{Fog6s} applications power usage}\label{fig:fog6s_applications}
	\end{minipage}

    \caption{Power usage of infrastructure and applications for different experiments.}
    \label{fig:test}
\end{figure*}

\subsubsection{Cloud Only}

In the first experiment, all processing tasks of both application types were placed in the cloud. There exist no fog nodes. 
Figures \ref{fig:cloud_infrastructure} and \ref{fig:cloud_applications} depict the consumption of infrastructure components and application types over time. We can observe that the major part of the power consumption, namely \SI{66.4}{\percent}, can be attributed to the WAN as all raw sensor data have to travel to the processing tasks placed in the cloud. 
The accumulated power required by V2I applications correlates with the number of taxis on the map. 

\subsubsection{Fog Only}

In experiment \emph{Fog 6}, six traffic light systems were equipped with fog nodes, to avoid data transfer over the power-intensive WAN. These nodes provide sufficient computing capacity so no tasks have to be offloaded to the cloud. Processing tasks are distributed evenly across the fog nodes. As expected, the power usage of network decreased sharply. The fog nodes themselves are now the most relevant power consumers, accounting for \SI{94.7}{\percent} of the total energy consumption. A major fraction of this, namely \SI{57.8}{\percent}, is static power consumption: The six fog nodes require \SI{100}{W} of static power each, totaling in \SI{14.4}{kW} during a day.

Although the computational and network load required by the CCTV applications is constant during the entire simulation, the reported power consumption varies over time. This is because we allocate the static power consumption of fog nodes proportionally to applications running on them and fog nodes are utilized inefficiently in this experiment. Especially at night when only a few taxis are on the road, the relative power demanded by the CCTV applications rises.

\subsubsection{Fog and Cloud}

Experiments \emph{Fog 1} to \emph{Fog 5} contain one to five fog nodes, respectively. 
Tasks are still distributed evenly across the available fog nodes, but once all fog nodes are running at more than \SI{85}{\percent} capacity, tasks are being offloaded to the cloud.
The lowest overall energy consumption was achieved in experiment \emph{Fog 4}, saving \SI{2300}{Wh} compared to \emph{Fog 6}. Four fog nodes provide enough capacity to host all CCTV processing tasks and around 125 V2I processing tasks. Although during some periods of the day more than 125 taxis are on the map simultaneously, the savings on static fog node power usage outweigh the additional cloud and WAN power usage caused by task offloading with four fog nodes.

\subsubsection{Energy-Saving Mechanism}

\begin{figure}[b!]
    \includegraphics[width=1\columnwidth, trim=0.8cm 1.5cm 1.8cm 2.14cm, clip]{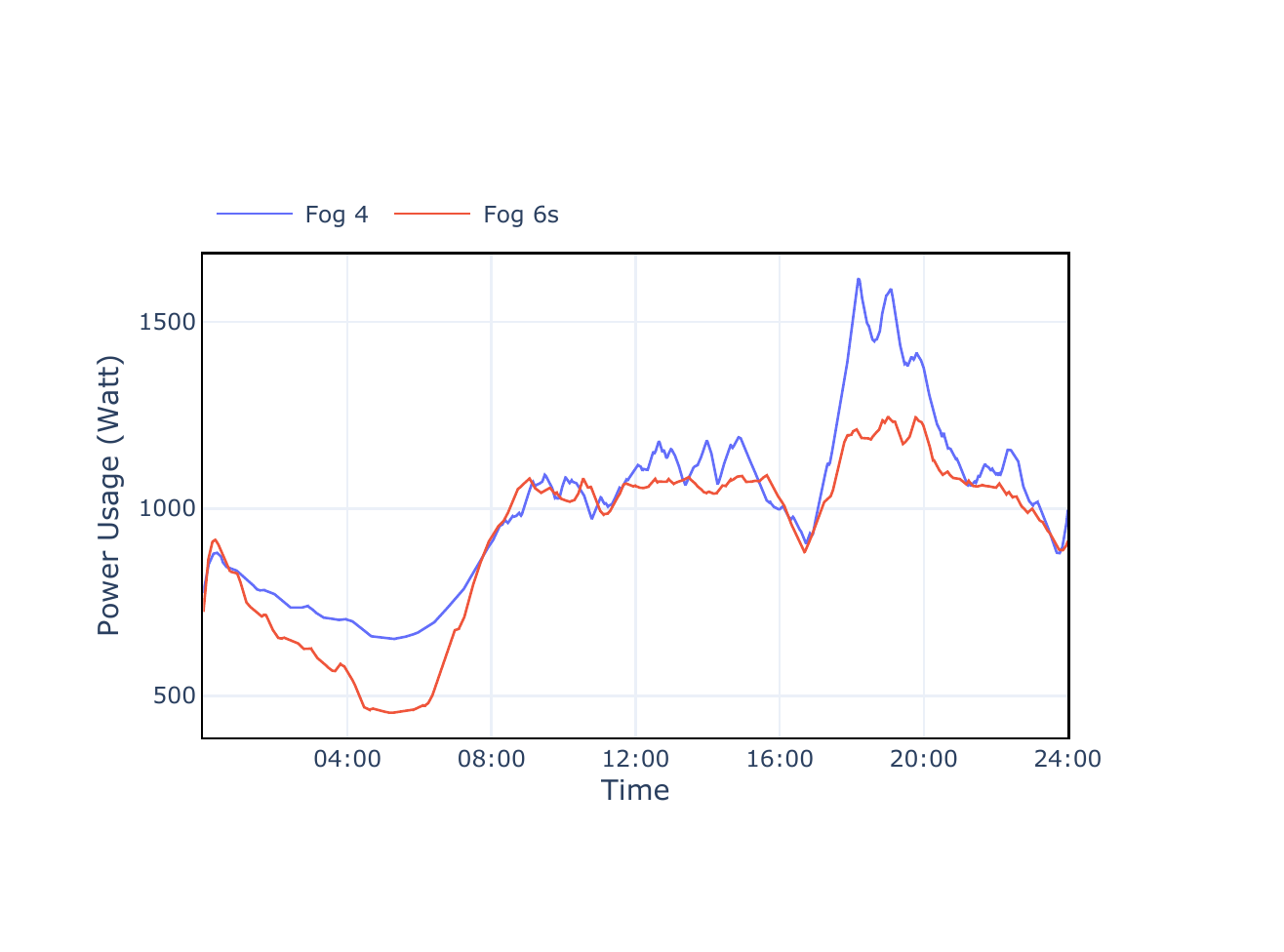}
    \caption{Comparison of overall power usage in experiments \emph{Fog~4} and \emph{Fog~6s} (smoothing applied).}
    \label{fig:fog4_vs_fog6s}
\end{figure}

Experiment \emph{Fog 6s} extends experiment \emph{Fog 6} with an adaptive task placement strategy and energy-saving mechanism.
This time, if a fog node is idle for five seconds, it will be put to sleep, reducing its static power consumption of zero. In order to fully exploit this energy-saving mechanism, the task placement algorithm tries to consolidate as much work as possible on a minimum number of nodes to maximize the number of idle fog nodes. Individual fog nodes are still only utilized up to \SI{85}{\percent} capacity.
Figure~\ref{fig:fog6s_infrastructure} shows that the overall static power usage of fog nodes is now no longer constant. All six fog nodes are only active for less than 2 hours during the day, while during the night only two out of six are active.
Consequently, the power usage of applications was reduced too, see Figure~\ref{fig:fog6s_applications}. 
Resulting from the improved placement strategy, CCTV applications are now responsible for less static power usage, and, thus, have a very stable load profile.
V2I applications benefit from the energy-saving mechanism by becoming more power proportional. At night, the attributed power usage of all V2I applications drops below \SI{80}{W}.

Figure~\ref{fig:fog4_vs_fog6s} displays how experiment \emph{Fog 6s} undercuts the second best performing experiment \emph{Fog 4} at different times of the day. When there is little traffic in the city, \emph{Fog 6s} benefits from highly reduced static power usage of shut-off fog nodes. In times of high utilization, \emph{Fog 6s} continues to provide sufficient resources to process all data in the fog layer, avoiding expensive WAN traffic for offloading. At times of average utilization, the load profile of both experiments is almost identical. %

\subsection{Analysis}

So far the evaluation has shown that we have successfully met our Requirements (1), (2), and (3): LEAF was used to model a realistic fog computing scenario with different kinds of compute nodes, network links and applications. By assigning individual power models to different parts of the infrastructure, we were able to analyze the energy footprint of different fog node deployments in order to find the most energy-conserving configuration. In the last experiment, we demonstrated how our model enables research on dynamic task placement strategies and energy-saving mechanisms.

To evaluate Requirement (4), \emph{Performance and Scalability}, we analyzed the runtimes of simulations. Each presented experiment simulates around \SI{46500}{} taxis executing around \SI{330000}{} tasks and finishes in less than 50 seconds on a single core of a \SI{1.4}{GHz} Intel Core i5.
A series of further experiments show that in the outlined scenario, the only algorithmic component that shows non-linear growth is the algorithm for finding shortest paths between tasks. Even when scaling experiments by factor 10, namely \SI{465000}{} taxis and \SI{3300000}{} tasks, they had a median runtime of 613 seconds.
Consequently, the presented model is suitable to simulate large-scale experiments several magnitudes faster than real time on commodity hardware.

\section{Conclusion}\label{sec:CONCLUSION}

This paper presents LEAF, a simulator for modeling large energy-aware fog computing environments.
LEAF features a holistic but granular energy consumption model that covers data centers, edge devices, and network as well as applications which are running on this infrastructure.
By combining analytical and discrete-event modeling, the proposed model enables the simulation of thousands of streaming applications on a distributed, heterogeneous, and dynamic infrastructure.
The publicly available implementation of LEAF was evaluated within a realistic smart city traffic scenario and proved to be a valuable tool for research on energy-conserving fog computing architectures, task placement strategies, and energy-saving mechanisms.

In the future, we plan to extend our model to provide time-based and location-based calculations of the carbon emissions and electricity costs, thus enabling research on infrastructures and algorithms optimized for a low environmental footprint and operating expenditure.
Moreover, we plan to integrate LEAF with relevant simulators for network, urban traffic, or electric grids to co-simulate and evaluate more realistic scenarios.

\section*{Acknowledgment}
This work was supported by the German Ministry for~Education and Research (BMBF) as BIFOLD (research grant 01IS18025A) and the German Academic Exchange Service (DAAD) as ide3a.

\balance

\bibliographystyle{IEEEtran}
\bibliography{bibliography}

\end{document}